\documentstyle[11pt,aaspp4]{article}

\begin{document}

\title{Lack of Evolution in the Iron Abundance in Clusters of Galaxies
and Implications for the Global Star Formation Rate at High Redshift}
\author{R. F. Mushotzky and M. Loewenstein\altaffilmark{1}}
\affil{Laboratory for
High Energy Astrophysics, NASA/GSFC, Code 662, Greenbelt, MD 20771}
\altaffiltext{1}{Also with the University of Maryland Department of Astronomy}

\begin{abstract}
We present the first large sample of accurate iron abundances 
and temperatures for clusters at redshifts $>0.14$. We find that 
the Fe abundance shows little or no evolution out 
to $z\sim 0.3$. This and the early formation epoch of 
elliptical galaxies in clusters
indicate that most of the enrichment of the intracluster medium 
occurred at $z>1$. If clusters represent fair samples of the universe,
then global metal production is 2-5 
times greater than is inferred from recent  
studies of galaxies at low and high
redshifts. 
\end{abstract}


\section{Introduction}

The hot 
intergalactic medium in 
low redshift clusters of galaxies has been enriched in iron to 
$\sim 1/3$ solar (\cite{e91}, \cite{y92}) by processes that, if
understood, are a 
powerful diagnostic of galaxy formation. Since
the mass in the x-ray emitting gas exceeds 
that in the stars in galaxies by factors of 2-10 (assuming
$H_o=50$ km s$^{-1}$ Mpc$^{-1}$ as we do throughout this
paper; we also assume $\Omega=1$), the IGM is the main 
reservoir of metals in clusters. 
Moreover, the strong correlation of iron 
mass with optical light (\cite{a92}) implies
a coupling of the constraints on the evolution of 
metals from the x-ray data on high redshift clusters and the 
optical data on the evolution of elliptical galaxies.  
If, as many theorists believe, clusters are fair 
samples of the universe, as well as the 
only systems likely to have retained all the metals produced by
stars, then
studies of the evolution of the 
IGM metallicity strongly constrain the
star formation history in the universe. 

There were few measurements of 
Fe in clusters at $z>0.1$
before the launch of the {\it ASCA} satellite 
(\cite{t94}). Neither early 
results from {\it ASCA} (\cite{mus96}, \cite{t96}), nor subsequent
detailed individual 
measurements (\cite{b94}, \cite{d96}, \cite{m96}), showed
evolution in the Fe abundance out to $z\sim 0.3$.
Here we report on a large sample of high signal-to-noise {\it ASCA} 
spectra of clusters at $z>0.14$, and 
find no evidence for evolution of the cluster metal
abundance. Since recent optical data  
reveals only passive evolution 
in cluster ellipticals out to $z\sim 0.4$,
most of the 
cluster metals were produced at 
$z>1$. We thus infer much higher star formation rates
in the $z>1$ universe than 
derived from optical/UV surveys.

\section{Observations}

We extracted and analyzed
$\sim 40$ spectra of clusters at $z>0.14$ from data in the {\it ASCA} 
public archives before 12/96.
We concentrate on the 21 high signal-to-noise 
observations with 
90\% confidence error in Fe abundance $<0.12$ --
a moderately large sample upon which
simple and unambiguous statistical tests can be made.
At $kT \sim 6$ keV,
the product of the flux, $f_x$, and exposure time must exceed
$\sim 1.4\ 10^{-7}$ erg cm$^{-2}$ to attain this accuracy. 
For a typical {\it ASCA} exposure time of 
40 ks (net exposures for two clusters in 
our sample, MS1358 and AC118, were $\sim 70$ ks), a 
minimum flux $f_{min}=3.5\ 10^{-12}$ erg cm$^{-2}$ s$^{-1}$
is thus required.
Before the release of the {\it ROSAT} all-sky survey, such objects
were rare: e.g., in the IPC database  
only 19 clusters have $z>0.14$ and $f_x>f_{min}$.
Cross-correlation of the {\it RASS} and
Abell catalogs yields 50 clusters with $f_x>1.4f_{min}$
at $z>0.14$ (\cite{e96}), but
further cross-correlation with the {\it ASCA} observation log identifies 
only 19 clusters with $f_x>f_{min}$.
Thus we include most of the high signal-to-noise,
high-$z$ clusters.
(The flux from one cluster that
apparently met our criterion, Abell 1774, turned 
out to be dominated by an AGN.) 
This subsample (Table 1) 
spans the bolometric x-ray luminosity 
(about twice the 2-10 keV 
luminosity) range from 
$10^{45.1}$ to $10^{46.3}$ erg s$^{-1}$, overlapping 
the low-$z$ sample measured by 
previous 
satellites. 

We applied standard 
processing 
to the data (e.g., \cite{mu96}); and,
for both 
the SIS and
GIS, used 
a $6'$ extraction radius, corresponding 
to 3 Mpc at $z\sim 0.3$ -- sufficiently extended to  
include the total
flux. At $z\sim 0.15$, 10-30\% of the flux may fall outside the 
extraction radius.
We used all-sky backgrounds accumulated during the {\it ASCA} deep 
survey pointings
for clusters with $z<0.2$, and
background from the same field for the higher $z$ systems.
Spectra were fit to an isothermal, collisional 
equilibrium plasma model at the 
optical redshift of the source. The redshifts
obtained from {\it ASCA} data 
are within 0.013 of the optical values
(e.g., for the relatively faint MS1358 the fitted redshift is
$z_{fit}=$ 0.323(+.011,-0.022) compared to 
$z_{opt}=$ 0.327, while for the brighter ZW3146
$z_{fit}=$ 0.292 (+0.004,-0.0061) and
$z_{opt}=$ 0.291).

Cooling-flow-plus-isothermal-plasma 
models were used
in cases (ZW3146, A2390) where a cooling flow is clearly 
present in the 
{\it ROSAT} image (cf. \cite{a96}).
As found by
\cite{a96},
the extra component has little effect on
the derived abundances of the
isothermal plasma.
Our fits
are in excellent agreement with results for those
clusters in Table 1 that were previously analyzed: RXJ 1347 (\cite{s96}),
ZW3146, A1835 and 
E1455 (\cite{a96}), A1204 (\cite{m96}), and MS0451 (\cite{d96}). (The
results are new for all others.)
The parameters derived from 
{\it ASCA} spectra (Figure 1) are robust for these high flux objects. 

\begin{deluxetable}{cccccc}
\footnotesize
\tablecaption{High Signal-to-Noise Clusters} 
\tablewidth{530pt}
\tablehead{
\colhead {Cluster} &
\colhead{log $L_x$\tablenotemark{a}} &
\colhead{log $kT$ (keV)} &
\colhead{$Z_{Fe}$\tablenotemark{b}} &
\colhead{$\delta Z_{Fe}$\tablenotemark{c}} &
\colhead{$z$\tablenotemark{d}} }
\startdata
A1204 & 45.08 &	0.587 & 0.350 &	0.07 & 0.170 \nl
A1722 & 45.23 &	0.768 & 0.250 & 0.11 & 0.270 \nl 
A1246 &	45.18 &	0.798 &	0.223 & 0.08 & 0.187 \nl
MS2137-24 & 45.25 & 0.643 & 0.410 & 0.12 & 0.313 \nl
A963 &	45.30 &	0.829 &	0.291 & 0.08 & 0.206 \nl
MS1358+62 & 45.30 & 0.813 & 0.272 & 0.10 & 0.327 \nl
A1413 &	45.36 &	0.797 &	0.289 &	0.05 & 0.143 \nl
A1763 &	45.38 &	0.953 &	0.263 &	0.09 & 0.187 \nl
A2218 &	45.41 & 0.857 & 0.180 &	0.065& 0.171 \nl 
A773 &	45.45 & 0.952 &	0.240 &	0.08 & 0.197 \nl
MS1455 & 45.47 & 0.660 & 0.330 &	0.08 & 0.258 \nl
A520 & 45.48 & 0.934 & 0.250 &	0.20 & 0.201 \nl
MS0451-03 & 45.60 & 1.01 & 0.157 & 0.12 & 0.539 \nl
AC118 &	45.65 & 0.970 &	0.228 & 0.09 & 0.308 \nl
A2390 &	45.73 &	0.949 &	0.220 &	0.06 & 0.230 \nl
ZW3146 & 45.74 & 0.802 & 0.240 & 0.05 & 0.290  \nl
A1689 &	45.77 & 0.947 &	0.260 & 0.06 & 0.180 \nl
A2204 &	45.78 &	0.903 &	0.450 &	0.05 & 0.153 \nl
A2219 &	45.92 &	1.07 &	0.250 & 0.07 & 0.228 \nl
A1835 &	45.92 &	0.910 &	0.320 & 0.05 & 0.252 \nl
RXJ1347-11 & 46.28 &  0.968 & 0.330 & 0.10 & 0.451 \nl

\enddata

\tablenotetext{a}{Log bolometric x-ray 
luminosity in erg s$^{-1}$ assuming $q_o=0$ and $H_o=$
50 km s$^{-1}$ Mpc$^{-1}$.}
\tablenotetext{b}{Fe 
abundance with respect to the \cite{a89} photospheric abundances.}
\tablenotetext{c}
{The error in the Fe abundance -- 90\% confidence for one parameter 
($\Delta\chi^2=2.7$).}
\tablenotetext{d} {Optically determined redshift.}

\end{deluxetable}

\section{Cluster Metallicity as a Function of Redshift}

Our sample of 21 clusters has nearly identical
Fe abundance
properties
to the low-$z$ {\it Ginga} sample
(\cite{y92}, \cite{b95}),
{\it i.e.} a 
mean abundance $\langle$Fe$\rangle=0.27\pm 0.013$ and variance 0.076
(Figure 2)
compared to 
$\langle$Fe$\rangle=0.29\pm 0.014$ and variance 0.062
(all errors are 90\% confidence for one parameter of interest).
A K-S
test shows no significant difference between the samples. 
We find no 
evidence of systematic abundance
variations with luminosity. 
The weak abundance-temperature correlation (Figure 3)
is not significant at 90\% confidence. 
We conclude that this sample, with $\langle$z$\rangle=0.26$, shows no 
statistical differences in Fe abundance from a similar sample of low-$z$ 
objects. Results for the 
systems with larger abundance errors agree with the above conclusions. 

Although our
sample is not flux- or redshift-limited, it has no obvious
selection biases and  
should be representative 
of  $z\sim 0.25$ clusters. Its mean luminosity, 
$\langle$log $L_{bol}\rangle=45.4$, and 
mean temperature, $\langle kT\rangle=7.5$ 
keV, are slightly higher than in the low-$z$ sample 
drawn from a flux limited 
survey ($\langle$log $L_{bol}\rangle=45.16$, $\langle kT\rangle=5.96$ keV).
The temperature-luminosity
relationships of the $z\sim 0.25$ and low-$z$ samples
show no significant differences
(\cite{m97}).

\section{Discussion}

In high-$\Omega$ hierarchical clustering  
scenarios (\cite{k95}), the mass of
rich clusters 
grows by $\sim 2$ from $z\sim 0.2$ to the present. Moreover,
systems at $z\sim 0.4$ evolve
differently from a cluster 
of the same richness observed today, with high-$z$ clusters 
assembled over a shorter time interval and 
undergoing more merging in the few Gyr prior to the epoch when
they are observed. Thus 
the present and low-$z$ samples should differ 
if the accreted material has distinct properties, and 
the surprising constancy of 
both the mean and variance of the Fe distribution 
over this redshift range (corresponding to a lookback time 
of $\sim$1/4-1/3 the age of the universe) argues for a very 
homogeneous accretion process or a low-$\Omega$ universe where 
cluster growth is essentially completed at $z\sim \Omega^{-1}$ 
(\cite{w96}). 
The substantial spread in cluster metallicity at 
$z=0$ and $z=0.25$ is unexpected
within purely gravity-driven hierarchical 
clustering scenarios. 
Many of these clusters include an
excess of blue galaxies compared to low-$z$
systems (the ``Butcher-Oemler effect''). None of these 
have anomalous metallicity: whatever 
processes converted high-$z$
blue spiral galaxies into low-$z$ spheroidal systems had 
no major impact on the IGM.

The absence of evolution over the past $\sim 5$ Gyr implies early
enrichment of the IGM. Thus,
ram pressure stripping of infalling galaxies at low-$z$
cannot dominate the enrichment, but 
much of the cluster Fe could be
produced by SNIa (\cite{r93}): 
reconciling large Fe-mass-to-light ratios with
modest present-day SNIa rates
requires the rate to have a short rise time followed by steep
negative evolution such that 90\% of the Fe is
produced at $z>0.5$.
Most of the cluster metals, perhaps including 
most of the Fe, were produced at high $z$ by SNII (\cite{l96}).

The stars in elliptical 
galaxies that we see today come from the same population that 
produced the metals in the IGM
(e.g., \cite{e95});
moreover, the elliptical 
galaxies in several rich $z\sim 0.3$-0.4 clusters of galaxies
show only 
passive evolution
(e.g., \cite{b96}). Therefore, the 
star formation in 
elliptical galaxies that produced the cluster metals 
must have stopped $>2$ Gyr before $z\sim 0.4$, placing
the period of cluster metal formation at $z>1$ -- in excellent
agreement with the lack of evolution inferred from {\it ASCA} data. 
This is in contrast to the history
of metal production by disk galaxies, where star formation peaked
at $1<z<2$ and continues through the present epoch
(\cite{p95}, \cite{mad96}). 

The containment by the 
deep cluster potential of the metals synthesized and ejected by
supernovae during the epoch when the stars that make up
present-day elliptical galaxies formed, enables us to calculate
the $z>1$ metal production by 
(proto-)ellipticals
and estimate their contribution
to the total metals produced at $z>1$ and
over the Hubble time. 
Assuming that relative abundances in clusters are constant,
the mass in metals produced per unit (present-day) blue-band
elliptical galaxy luminosity is $\sim 0.3$ in solar units.
If the star formation (though not necessarily the {\it dynamical} evolution)
in elliptical galaxy precursors 
proceeded in a similar way in the field and in clusters, then
the product of the metal mass-to-light ratio and the 
early-type galaxy luminosity
density (\cite{l92}) yields the total
mass density of metals from early-type galaxies,

\begin{equation}
\rho_{z}=1.4\ 10^7 M_{\odot} Mpc^{-3}.
\end{equation}

\noindent
This is consistent
with the value, derived by \cite{z96}, required to
account for the increase in
stellar mass-to-light ratio with luminosity as due to
stellar remnants from primordial star formation with an IMF biased
towards high mass stars.

The metals ejected by proto-ellipticals, as
calculated above, exceed what is estimated to be locked
inside the stars in normal massive galaxies today 
by 2-5 (\cite{mad96}). Since the latter is
consistent with the star formation rate derived from galactic
UV light, star formation in proto-ellipticals must have
proceeded in an obscuring environment or in sub-luminous fragments
(\cite{l96}). Since half the star formation in disks
has occurred since $z=1$ but, 
star formation in ellipticals was completed by $z=1$
(cf. \cite{c96}),
proto-spheroids increasingly dominate
star formation
at early epochs. If elliptical galaxy
star formation
predominantly occurred at $1<z<6$, the average metal production
rate was

\begin{equation}
{{d\rho_{z}}\over {dt}}=3.4\ 10^{-3} M_{\odot} yr^{-1} Mpc^{-3},
\end{equation}

\noindent
compared to averages from high-$z$, UV surveys of galaxies
of 0.89, 0.46, and 0.17 
$\times 10^{-3} M_{\odot}$ yr$^{-1}$ Mpc$^{-3}$ for $z>1$, $z>2$, and $z>3$,
respectively.

Thus, global star formation rates
derived from UV surveys are severe underestimates.
At $z>0.3$, where $H\alpha$ redshifts out of the 
ground-based optical band, [OII]3727 or 
the UV continuum are used as star formation indicators. 
Recent ASTRO-2 UIT and HUT observations (\cite{fa96}) show
that there is often 
a shift of a factor 2 in specific star formation rates using
the UV continuum instead of $H\alpha$; and
furthermore, comparison of IRAS $60\mu$ luminosities with 
$H\alpha$ values indicates another factor of 2 offset
(see also,
\cite{b92}, \cite{d94}). Since 
even small amounts of reddening are sufficient to remove 
objects from the high-$z$ galaxy samples and to change their 
inferred star formation rates, we believe that the cluster data 
are much more robust indicators of the true metal 
formation history of the universe.        

\section{Conclusions}

The Fe abundance of the
intergalactic medium in rich clusters of galaxies
undergoes little if any 
evolution from $z\sim 0.3$ to $z\sim 0$. This strongly constrains  
high-$\Omega$ hierarchical theories in which clusters rapidly grow 
during this period. A real and constant
spread in the metallicity at all observed redshifts implies
stochasticity, not only in the construction of these massive
systems, but also in the efficiency and/or timing of star formation
within their constituent galaxies.

A self-consistent history of star formation in 
disk galaxies has emerged from a synthesis
of  observations of damped $L\alpha$ systems, deep ground based and
{\it Hubble Space Telescope} surveys of star forming galaxies,
and spectral synthesis and chemical evolution modeling.
However, this census of star
formation in the universe is incomplete, since
most of the metals produced at high $z$ were expelled from galaxies.
A complementary picture of star formation
in proto-spheroids is emerging, driven by  
the lack of evolution in IGM metallicity reported here
and optical evidence
that together indicate that
cluster metals originate from the precursors of elliptical galaxies
at $z>1$. 
The $L$-dependence of $M/L$ (\cite{z96}) and 
gross and relative IGM elemental abundances (\cite{l96}) are explained
if this star formation produces
an IMF skewed toward high masses, 
precipitating a massive outflow of metal-enriched gas. This star
formation must occur in sub-luminous fragments or be dust-enshrouded
to avoid detection by UV surveys, but the IR and/or UV
background produced is within reach of near-future measurements (\cite{z96}).
The IGM of rich clusters are reservoirs for primordial metal-rich
galactic outflows driven by this
star formation, and their investigation with
{\it ASCA} has enabled us to discover that
$\sim 4$ times more metals were created at $z>1$ than 
are inferred from high redshift galaxy data. 

\acknowledgments

We thank Prof. Y. Tanaka for early encouragement of this work, and
acknowledge the efforts of the entire {\it ASCA} team.

\clearpage

\figcaption{68, 90 and 99\% error contours for the temperature 
and Fe abundance in three rich clusters, as indicated,
illustrating clusters with small, typical, and large
abundance uncertainties.}

\figcaption{Fe abundance {\it vs} redshift.
The error bars show 90\% confidence uncertainty limits.}

\figcaption{Fe abundance {\it vs} gas temperature}


\clearpage

\begin{thebibliography}{}
\bibitem[Anders \& Grevesse 1989]{a89}
Anders, E., \& Grevesse , N. 1989, Geochimica et Cosmochimica Acta, 53, 197
\bibitem[Allen et al. 1996]{a96}
Allen, S. W., Fabian, A. C., Edge, A. C., Bautz, M. W., Furuzawa, A.,
\& Tawara, Y. 1996, MNRAS, 283, 263
\bibitem[Arnaud et al. 1992]{a92}
Arnaud, M., Rothenflug, R., Boulade, O., Vigroux, L., \& Vangioni-Flan, E. 
1992, A\& A, 254, 49
\bibitem[Bautz et al. 1994]{b94}
Bautz, M. W., Mushotzky, R., Fabian, A. C., Yamashita, K.,
Gendreau, K. C., Arnaud, K. A., Crew, G. B., \& Tawara, Y. 1994, 
PASJ, 46, L131 
\bibitem[Bender et al. 1996]{b96}
Bender, R., Ziegler, B., \& Bruzual, G. 1996, ApJ, 463, L51
\bibitem[Buat 1992]{b92}
Buat, V. 1992, A\&A, 264, 444
\bibitem[Butcher 1995]{b95}
Butcher, J. 1995, Ph.D. Thesis, Leicester University
\bibitem[Cowie et al. 1996]{c96}
Cowie, L., Songaila, A., Hu, E. M., \& Cohen, J. G. 1996, AJ, 112, 839
\bibitem[Deharveng et al. 1994]{d94}
Deharveng, J.-M., Sasseen, T. P., Buat, V., Bowyer, S., Lampton, M,
\& Wu, X. 1994, A\&A, 289, 715
\bibitem[Donahue 1996]{d96}
Donahue, M. 1996, ApJ, 468, 79
\bibitem[Ebeling et al. 1996]{e96}
Ebeling, H., Voges, W., Bohringer, H., Edge, A. C., Huchra, J. P.,
Briel, U. G. 1996 MNRAS, 281, 799
\bibitem[Edge \& Stewart 1991]{e91}
Edge, A., \& Stewart, G. 1991, MNRAS, 252, 428
\bibitem[Elbaz et al. 1995]{e95}
Elbaz, D., Arnaud, M., \& Vangioni-Flam, E. 1995, A\&A, 303, 345
\bibitem[Fanelli et al. 1996]{fa96}
Fanelli, M. et al. 1996, preprint (astro-ph/9612086)
\bibitem[Kauffmann 1995]{k95}
Kauffmann, G. 1995, MNRAS, 274, 153
\bibitem[Loewenstein \& Mushotzky 1996]{l96}
Loewenstein, M., \& Mushotzky, R. F. 1996, ApJ. 466, 695
\bibitem[Loveday et al. 1992]{l92}
Loveday, J., Peterson, B. A., Efstathiou, G., \& Maddox, S. J. 1992,
ApJ, 390, 338
\bibitem[Madau et al. 1996]{mad96}
Madau, P., Ferguson, H. C., Dickinson, M.,
Giavalisco, M., Steidel, C. C., \& Fruchter, A. 1996, preprint
(astro-ph/9607172)
\bibitem[Matsuura et al. 1996]{m96}
Matsuura, M., Miyoshi, S. J., Yamashita, K., Tawara, Y., Furuzawa, A.,
Lasenby, A. N., Sanders, R., Jones, M., \& Hatsukade, I. 1996, ApJ, 466, L75 
\bibitem[Mushotzky 1996]{mus96}
Mushotzky, R. F. 1996, in Roentgenstrahlung from the Universe, ed.
H. U. Zimmerman, J. Trumper, and H. Yorke (MPE Report 263), 545
\bibitem[Mushotzky et al. 1996]{mu96}
Mushotzky, R. F., Loewenstein, M., Arnaud, K. A., Tamura, T., Fukazawa, Y.,
Matsushita, K., \& Kikuchi, K. 1996, ApJ, 466, 686
\bibitem[Mushotzky \& Scharf 1997]{m97}
Mushotzky, R. F., \& Scharf, C. A. 1997, ApJ, submitted
\bibitem[Pei \& Fall 1995]{p95}
Pei, Y. C., \& Fall, S. M. 1995, ApJ, 454, 69
\bibitem[Renzini et al. 1993]{r93}
Renzini, A., Ciotti, L., D'Ercole, \& Pelligrini, S. 1993, ApJ, 419, 52
\bibitem[Schindler et al. 1996]{s96}
Schindler, S., Hattori, M., Neumann, D. M., Bohringer, H. 1996, A\&A,
submitted
\bibitem[Tanaka et al. 1994]{t94}
Tanaka, Y., Inoue, H., \& Holt, S. S. 1994, PASJ, 46, L37
\bibitem[Tsuru et al. 1996]{t96}
Tsuru, T., Koyama, K., Hughes, J. P., Arimoto, N., Kii, T., \& Hattori, M.
1996, in UV and X-ray Spectroscopy of Astrophysical and
Laboratory Plasmas, ed. K. Yamashita and T. Watanabe (Tokyo: Universal
Academy Press), 375
\bibitem[White 1996]{w96}
White, S. D. 1996, preprint (astro-ph/9608044)
\bibitem[Yamashita 1992]{y92}
Yamashita, K. 1992, in Frontiers of X-ray Astronomy, ed. Y. Tanaka and
K. Koyama (Tokyo: Universal
Academy Press), 475
\bibitem[Zepf \& Silk 1996]{z96}
Zepf, S. E., \& Silk, J. 1996, ApJ. 466, 114
\end{thebibliography}
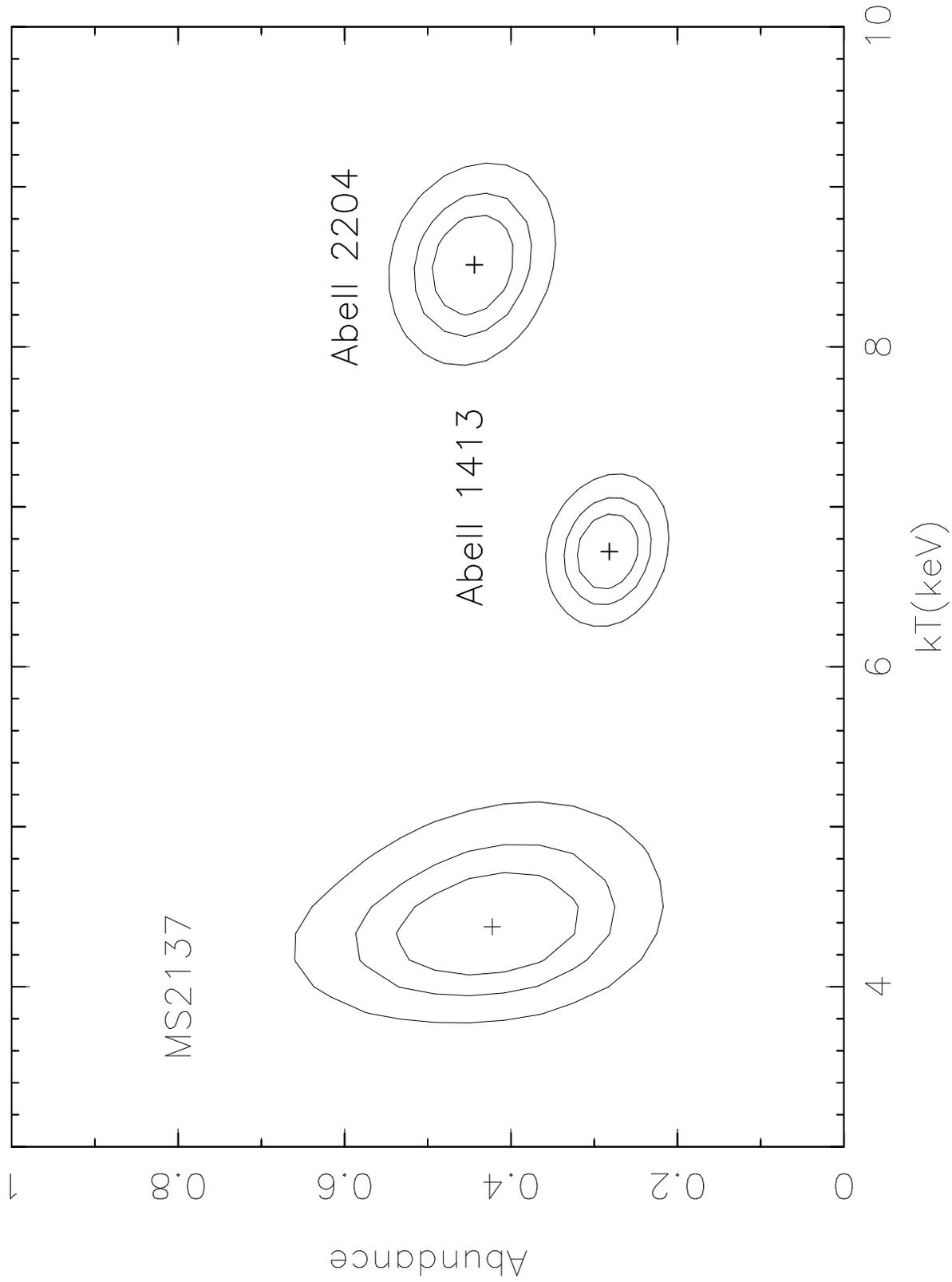
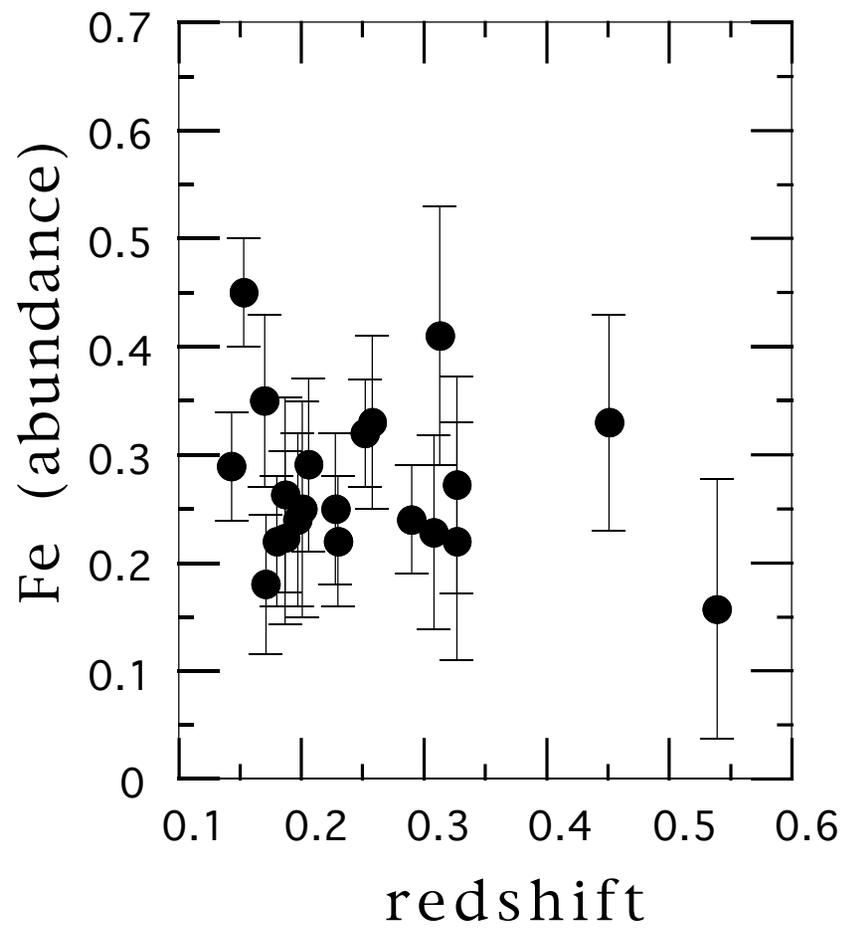
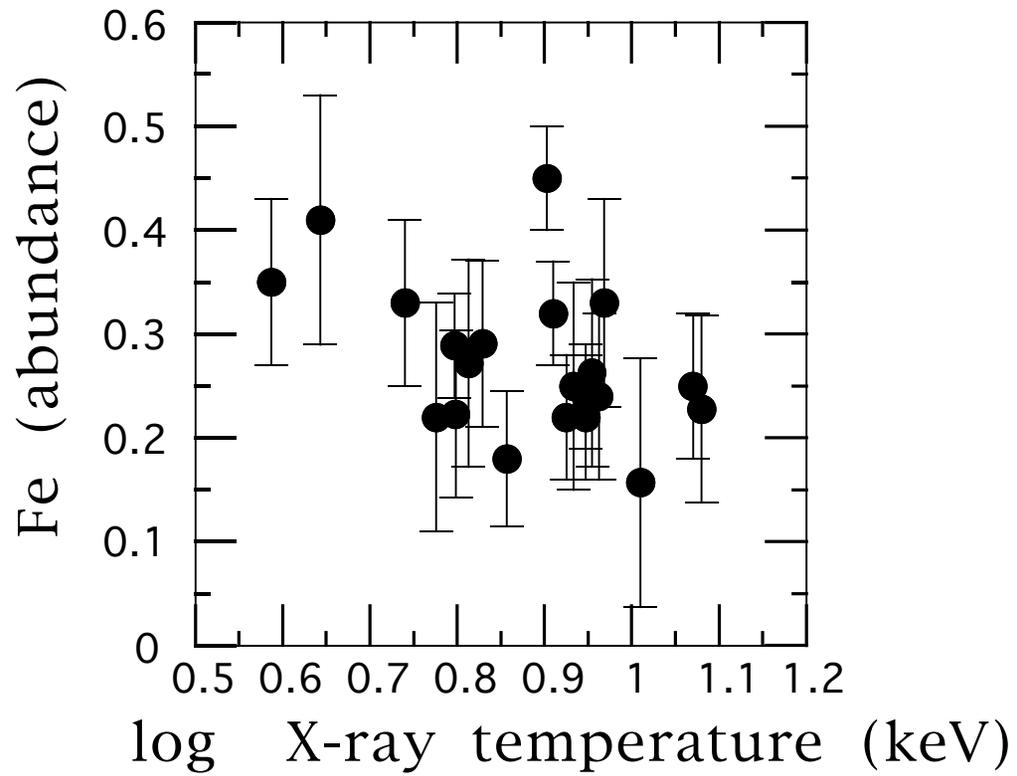
\end{document}